\documentstyle[twoside,fleqn,npb,epsfig]{article}
\newcommand{\AmS}{{\protect\the\textfont2
  A\kern-.1667em\lower.5ex\hbox{M}\kern-.125emS}}

\def\Bv{\not{\hbox{\kern-4pt $B$}}}
\def\Lv{\not{\hbox{\kern-4pt $L$}}}

\def \rpv{{R\hspace{-0.22cm}/}_P}

\def\be{\begin{equation}}
\def\ee{\end{equation}}
\def\bea{\begin{eqnarray}}
\def\eea{\end{eqnarray}}

\def\n{\noindent}

\def \zbs{{Z\to b\bar{s}}}

\def \st{\tilde{t}}
\def \sc{\tilde{c}}
\def \sb{\tilde{b}}
\def \ss{\tilde{s}}

\def \sq{\tilde{q}}
\def \slep{\tilde{\ell}}

\def \snu{\tilde{\nu}}

\def\gsim{\lower0.5ex\hbox{$\:\buildrel >\over\sim\:$}} 
\def\lsim{\lower0.5ex\hbox{$\:\buildrel <\over\sim\:$}}

\title{Rare $Z$ Decays
\thanks{Talk presented at RADCOR and Loops and Legs in Quantum Field
Theory 2002.}
}

\author{Gad Eilam\address{Department of Physics, Technion--Israel 
        Institute of Technology, \\
         32000, Haifa, Israel}
        \thanks{Based on work done in collaboration with D. Atwood,
        S. Bar-Shalom and A. Soni \cite{ourZbs}.}}
                
\begin{document}

\begin{abstract}
Motivated by the well known impact of rare decays of hadrons
and leptons on the evolution
of the Standard Model and on limits for new physics, as well 
as by the proposal for Giga-$Z$ option at TESLA, we investigate
the rare decay $Z \to b {\bar s}$ in various extensions of the 
Standard Model.  
\end{abstract}

\maketitle

\section{INTRODUCTION}

The central role played by rare decays on our understanding of
elementary particle physics, is well known, where ``rare''
stands here for Flavor Changing Neutral Currents (FCNC), which are
either small or practically vanishing in the SM. Some highlights:
are:\\
1. In $K$ physics: The first appearance of charm in loops from
which $m_c\approx 1.5$ GeV was predicted \cite{Gaill73}.\\
2. In $B$ physics: The importance of $b \to s\gamma$
in the SM and for extracting limits
on Beyond the SM (BSM) scenarios \cite{Lunghi}.\\
3. The top quark FCNC provide an excellent tool to 
investigate various extensions of the SM \cite{toprare}.\\
4. The experimental upper limit of the decay $\mu \to e\gamma$ 
\cite{mutoegam}, places
severe limits on extensions of the SM.
 
In view of the above and prompted by the recent discussion of a 
Giga-$Z$ option at TESLA \cite{GigaZ} in which the center-of-mass energy will 
be lowered to $M_Z$, producing more than $10^9$ $Z$ bosons ({\it i.e.}
$\sim 100$ times the number produced at LEP), one should
investigate rare decays of $Z$s. Now since the important subject
of rare leptonic $Z\to \ell_I {\bar \ell}_J,~\ell_I\neq\ell_J$ decays,
for which the SM branching ratio is $\leq 10^{-54}$, was covered
by Illana \cite{Illana}, we concentrate here on purely hadronic FCNC
$Z \to d_I {\bar d}_J,~d_I\neq d_J$ decays. In fact we only discuss
$Z \to b{\bar s}$ which in most models, including the SM, has the
largest branching ratio among hadronic FCNC $Z$ decays. Note however
that experimentally, the latter is practically inseparable from 
the $b{\bar d}$ mode. Let us also note that when referring 
to the $b{\bar s}$ mode, we actually mean $Z \to b{\bar s}+{\bar b}s$.
We note here that in the SM \cite{SMzbs} 
${\rm BR}(\zbs)\approx 3\times 10^{-8}$. 
 
In the following sections we will discuss two variants
of 2 Higgs Doublet Models (2HDM) and two of Supersymmetry (SUSY).
Of the latter the first one will be: SUSY with squark mixing,
while in the second one FCNC will result from SUSY with R Parity Violation
(denoted by RPV, or $\rpv$). 
As we will see, ${\rm BR}(\zbs)$
can be either smaller, the same or above the SM with
a maximal value of ${\rm BR}(\zbs)\approx 10^{-6}$.
 
Experimentally, the attention devoted to FCNC in hadronic $Z$ decays
at LEP and SLD has been close to nil. 
The best upper limit is \cite{DELPHI}
$\sum_{q=d,s}\rm{BR}(Z\to b {\bar q})\leq 1.8 \times 10^{-3}~@
~90\%~CL$.
This is a preliminary DELPHI limit (which will probably remain as
such forever...)
based on about $3.5\times 10^6$ hadronic decays. Experimentalists
who are privy to LEP and SLD data should be
encouraged to look in their data and improve the above limit. 

Due to space limitations, the following discussion of various
BSM models and their predictions for ${\rm Br}(\zbs)$, will be 
sketchy. Many more details and a more complete set of references
can be found in \cite{ourZbs}. In fact, almost each reference
should start with: ``See {\it e.g.}$\dots$'' and end with:
``$\dots$ and references therein.''

\section{GENERIC CALCULATION}

\begin{figure}
\hspace*{0.7cm}
 \includegraphics[height=9cm,angle=0]{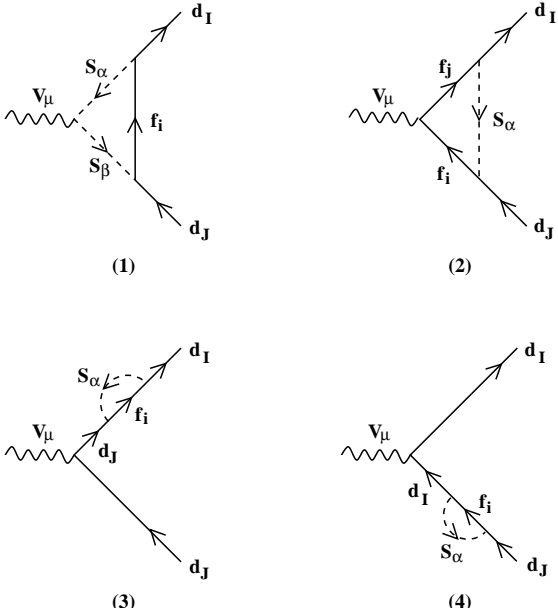}\\   
\caption{\label{fig1} One-loop diagrams that contribute to the
flavor changing transition $V \to d_I \bar d_J$, due to scalar-fermion
exchanges.}
\end{figure}

We start with a generic calculation of the diagrams which modify (at
one loop) the $V d_I {\bar d}_J$ vertex, due to charged or neutral scalar,
as depicted in Fig. \ref{fig1}.
In our case $V=Z$, $d_I=b$ 
and ${\bar d}_J={\bar s}$.  
The indices $i,j$ and $\alpha,\beta$ indicate
which fermions and scalars we are considering, 
respectively.
The Feynman rules are:\\
$V_\mu f_I f_J:~~i \gamma_\mu \left( a_{L(Vf)}^{ij} L +
 a_{R(Vf)}^{ij} R \right)$\\
$V_\mu S_\alpha S_\beta:~~i g_V^{\alpha \beta} 
\left( p_\alpha - p_\beta \right)_\mu$\\
$S_\alpha d_J f_i:~~i \left(
b_{L(\alpha)}^{ij} L + b_{R(\alpha)}^{ij} R \right)$,\\
where $L(R)=[1-(+)\gamma_5/]2$.\\
There are 4 one-loop amplitudes, each corresponding
to one of the 4 one-loop diagrams. Each amplitude is proportional 
to $\epsilon_\mu (q)$ times\\
$\bar u(p_b) 
\left[ \gamma^\mu \left( A_L L + A_R R \right) +
\left( B_L L + B_R R \right) p_\mu \right] v(p_s).$ $A_{L,R},~B_{L,R}$ are 
momentum dependent form factors, calculable from the diagrams.
There are 4 per diagram, thus we have 16 form factors. 
$A_L$ for diagram (1) is: \\
$A_L = -2 \sum_{\alpha,\beta,i} 
g_Z^{\alpha \beta} b_{L(\alpha)}^{iI} 
b_{L(\beta)}^{iJ} C_{24}$,\\
and similarly for the other 15 form factors.
$C_{24}$ is one of the usual one-loop scalar functions \cite{Pas:Velt} at 
$m_{f_i}^2,m_{S_\alpha}^2,m_{S_\beta}^2,
m_{d_I}^2,q^2,m_{d_J}^2.$\\
Finally:
\begin{eqnarray*}
\Gamma(Z\to b \bar s) = 
2\frac{N_C}{3} \left( \frac{1}{16 \pi^2} \right)^2 
\frac{M_Z}{16 \pi} \times
\left[ 2 \left( \mid A_L^T \mid^2
\right.\right. \\
\left.\left.  
+ \mid A_R^T \mid^2 \right)
+\frac{M_Z^2}{4}  \left( \mid B_L^T \mid^2 + \mid B_R^T \mid^2 \right) 
\right] ,
\end{eqnarray*}
where $A_L^T$ is the Total sum of $A_L$s from the 4
diagrams, and similarly for $A_R^T$, $B_L^T$ and $B_R^T$.

\section{MODELS AND PREDICTIONS}

The stage is now ready for identifying, for each 
model, the relevant scalars $S_\alpha$, fermions $f_i$ and the
couplings $a,b~\rm{and}~g$ (with the appropriate indices),
as expressed in the Feynman rules above. Then, the route
for obtaining $\Gamma(\zbs)$ using the generic equation in
the previous section is clear.
 
\subsection{Two Higgs doublet models}

In 2HDM with flavor diagonal couplings of the neutral Higgs 
to down-quarks, the FCNC $\zbs$ go
through the one-loop 
diagrams in Fig. \ref{fig1}. The scalars are the charged Higgs bosons,
$S_{\alpha=1}=H^+$ and the fermions are $f_i=u_i,~i=1,2,3$. 
The couplings are:  
$Z_\mu u_i \bar u_j$ is as in the SM (therefore only $i=j$ survives), 
$Z_\mu H^+ H^-$ is derived from the kinetic energy part of the
Lagrangian ${\cal L}$ and $H^+ \bar u_i d_j$ is obtained from the 
Yukawa part which, in common notation is \cite{our_rev}:
\begin{eqnarray*} 
{\cal L}_Y &=& - \sum_{i,j}  \bar Q_L^i 
\left[ \left(U_{ij}^1 \tilde\Phi_1  +U_{ij}^2 \tilde\Phi_2 \right) u^j_R 
\right. \\
&& \left. + 
\left(D_{ij}^1 \Phi_1  +D_{ij}^2 \Phi_2 \right) d^j_R \right].
\end{eqnarray*}
A choice of $U$ and $D$, which are 
$3 \times 3$ matrices in flavor space, leads to
a specific 2HDM. We now study two variants of 2HDM.
 
\subsubsection{Two Higgs doublet model of type II}
In this model, called 2HDMII, $U^1=D^1=0$, $\zbs$ was considered before
\cite{our4}. Using realistic values in the $\tan{\beta}-m_{H^+}$ plane,
we obtain: $\rm{BR}(\zbs)\lsim 10^{-10}$, two orders of magnitude below 
the SM.

\subsubsection{Two Higgs doublet model ``for top''}
In this variant \cite{our17:19}, named T2HDM, the top is rewarded for
its ``fatness'' by having its mass proportional to the large $v_2$,
while all other masses are proportional to $v_1$. It therefore makes
sense to consider here only $\tan{\beta}>>1$. Using T2HDM parameters 
consistent with data we find: $\rm{BR}(\zbs)\lsim 10^{-8}$.

\subsection{Supersymmetry with squark mixing}

FCNC in SUSY can emanate from squark mixing in:

\begin{eqnarray*}
&{\cal L}_{\rm{soft}}^{\rm {squark}}&=
-\tilde Q_i^\dagger (M_Q^2)_{ij} \tilde Q_j
-\tilde U_i^\dagger (M_U^2)_{ij} \tilde U_j\\
&-&\hspace{-0.5em}\!\!\!\!\!\!\!\!\tilde D_i^\dagger (M_D^2)_{ij} \tilde D_j
+A_u^{ij} \tilde Q_i H_u \tilde U_j+A_d^{ij} \tilde Q_i H_d \tilde D_j 
~, 
\end{eqnarray*}

\n with the usual notation for the squark fields \cite{ourZbs} 
and where $i,j$ are generation indices. Furthermore,

$M_{U, D}^2 = 
\left(
\begin{array}{cc}
(m_{\tilde U,\tilde D}^2)_{LL} & (m_{\tilde U,\tilde D}^2)_{LR} \\
(m_{\tilde U,\tilde D}^2)_{LR}^\dagger & (m_{\tilde U,\tilde D}^2)_{RR}
\end{array}
\right),$\\
\noindent where 
$(m_{\tilde U,\tilde D}^2)_{LL}$ ... 
are $3\times 3$ matrices. 
Under certain assumptions \cite{our20} and taking only
$\sb-\ss$ or $\st-\sc$ mixing into account:

\begin{eqnarray*}
(m_{\tilde U,\tilde D}^2)_{LL,RR}=
 \pmatrix{
1 &   0   &   0 \cr
0 &   1   &   \delta^{U,D(23)}_{LL,RR}  \cr
0 &  \delta^{U,D(32)}_{LL,RR} &  1 } m_0^2 ~.
\end{eqnarray*}

The above $\delta$s represent squark mixing from non-diagonal
bilinears in $\cal L$. $m_0$ is a common squarks mass scale, obeying:
$m_0>>M_Z$. Also,
$\delta_{LR}$s will stand for squark mixing from non-diagonal
trilinears in $\cal L$ \cite{ourZbs}. 
For them we adopt the Ansatz of \cite{our21},
leading to  $\delta_{LR}\propto vA/m_0^2$, 
where $A$ is a common trilinear soft breaking parameter for both up 
and down squarks.
$M^2_{D,U}$ become $4\times 4$ matrices in the weak bases
$\Phi^0_{D,U}=
(\ss_L,\ss_R,\sb_L,\sb_R),~(\sc_L,\sc_R,\st_L,\st_R).$
They are diagonalized to obtain the mass eigenstates
$\Phi_{D,U}=
(\ss_1,\ss_2,\sb_1,\sb_2),~(\sc_1,\sc_2,\st_1,\st_2),$
with the help of $R_{U,D}$ which rotates
$\Phi$ to $\Phi^0$.

We can now describe two cases of squark mixing: $\sb-\ss$
and $\st-\sc$ mixing.

\subsubsection{${\tilde b}-{\tilde s}$ mixing}
The scalars here are $S_\alpha=\Phi_{D,\alpha},~\alpha=1,2,3,4$, 
since $\sb-\ss$ admixture states run in the loops. 
The gluon is the only fermion in the loops, thus $f_i={\tilde g}$. The $a$
couplings are $0$, since $Z{\tilde g}{\tilde g}=0$. In other words,
one diagram (out of the four generic diagrams) vanishes.
The $b$ and $g$ couplings are functions of elements of the rotation
matrix $R_D$ mentioned above \cite{our22}. Since the two $\delta_{LR}
\lsim 10^{-2}$ \cite{our23}, we neglect them. For the other four 
$\delta$s we assume a common value, {\it i.e.}
$\delta^{D(23)}_{LL}=\delta^{D(32)}_{LL}=\delta^{D(23)}_{RR}=
\delta^{D(32)}_{RR}=\delta^D$, and vary
$0 < \delta^D < 1$.    

The parameters needed for masses, mixing and
$\Gamma(Z\to b \bar s)$ are:
$m_0,~\mu,~A,~\tan\beta,~m_{\tilde g}$ and $\delta^D$.
We vary them subject to $m_{\rm squarks}> 150$ GeV and
have plots of practically everything as a function of everything \cite{ourZbs}.
 
We find $\rm{BR}(\zbs)\lsim 10^{-6}$, where the 
highest value is attained for $m_{\tilde g}$ and one $m_{\tilde d_i}$
$\approx$ the EW scale,
while $m_{\tilde d_j}$, $j\ne i$ $\approx$ few TeV.
Such splitting
requires ``heavy'' SUSY mass scale with soft breaking parameters, 
which is consistent with the non-observability of SUSY particles so far. 

\subsubsection{${\tilde t}-{\tilde c}$ mixing}
In this scenario the 
scalars are $S_\alpha=\Phi_{U,\alpha},~\alpha=1,2,3,4$, similarly
to the previous case, except for $D\to U$. Obviously, 
$\st-\sc$ admixture states run in the loops. The loop fermions 
are the two charginos $f_i=\chi_i,~i=1,2$, and all 
four generic diagrams contribute to $\zbs$. The Feynman rules
\cite{our22} involve elements of the rotation matrix $R_U$
mentioned above and the chargino mixing matrices.

At the end of the day, running with the parameters
over all values consistent with the data, and with $m_{\tilde q}
>150$ GeV and $m_\chi>100$ GeV we obtain:
$\rm{BR}(\zbs)\lsim 10^{-8}$, which we could have anticipated since
BR(through $\st-\sc$ mixing): BR(through $\sb-\ss$ mixing) 
$\approx \alpha:\alpha_s$. 

\subsection{Supersymmetry with RPV}

Since there is no sacred principle which guarantees R-parity
conservation, we assume 
in this part of the talk that $R_P$ is violated. Then,
$\rpv$ terms in the SUSY superpotential $\cal W$
lead to FCNC. $\lambda$ terms (pure $\Lv$) in 
$\cal W$ are irrelevant at the 1-loop level. In addition 
we assume, for the pure $\Bv$ terms, that
$\lambda^{\prime \prime} << \lambda^\prime$, and also neglect the
bilinear term in the $\rpv$ part of $\cal W$. 
Then:
${\cal W}_{RPV} = \epsilon_{ab}
\lambda_{ijk}^\prime {\hat L}^a_i {\hat Q}^b_j {\hat D}_k^c.$
In addition, if 
$\rpv$ is OK then the $R_P$ conserving soft SUSY breaking is
extended. We need only the bilinear:
$V_{RPV} = \epsilon_{ab} b_i {\tilde L}^a_i H_u^b$,
where
$\tilde L$, $H_u$ are the scalar components of the hatted $L$ and 
$H_u$, respectively.
We therefore have two types of FCNC:\\
{\bf Type A:} Trilinear-trilinear: 
$\Gamma(\zbs) \propto (\lambda^\prime \lambda^\prime)^2$.\\
{\bf Type B:} Trilinear-bilinear:  
$\Gamma(\zbs) \propto (b \lambda^\prime)^2$.

\subsubsection{Type A: Trilinear-trilinear terms}
We further sub-divide type A contributions to 6 groups according to the 
scalars and fermions running in the loops. For instance, in
type A1 the scalars are $S_\alpha={\tilde e}_
{L,\alpha},~\alpha=1,2,3$ and the fermions are $f_i=u_i,~i=1,2,3$.
The $a$ couplings are identical to their SM values, $b_L=0$ (for
all $i,j$ and $\alpha$), $b_{R(\alpha)}^{i,j}=-\lambda_{\alpha i j}^
{\prime *}$ and $g_Z^{\alpha \beta}=-e(c^2_W-s^2_W)\delta_{\alpha\beta}/
2 s_W c_W$. Unfortunately, going over all type A groups, taking into
account the available limits on $\lambda{^\prime}$s and on
the other relevant parameters, we obtain
for the trilinear-trilinear case: $\rm{BR}(\zbs)\lsim 10^{-10}$.
Our results are in agreement with the special cases in \cite{our6}.

\subsubsection{Type B: Trilinear-bilinear terms}
In this case, a Higgs exchanged in the loop mixes with a slepton, through
$\epsilon_{ab} b_3 {\tilde L}^a_3 H_u^b$, 
assuming that only $b_3 \ne 0$.
We choose to work in the ``no VEV'' basis
$v_3=0$ in which:
$H_{u} \equiv \left(h_u^+,~ (\xi_{u}^0 + v_{u} + 
i \phi_{u}^0)/\sqrt{2} \right),$
$H_{d} \equiv \left((\xi_{d}^0 + v_{d} + i \phi_{d}^0)/\sqrt{2},~
h_d^- \right),$
$\tilde L_3 \equiv \left((\tilde\nu_{+}^0 + i
\tilde\nu_{-}^0)/\sqrt{2}~, \tilde e_{3}^- \right),$
where $\tilde\nu_{+}^0$, $\tilde\nu_{-}^0$, 
$\tilde e_{3}^-$ are 
SU(2) CP-even, CP-odd $\tau$-sneutrinos,
${\tilde \tau}_L$, respectively.
In the basis $\Phi_C^0 = (h_u^+,h_d^+,\tilde e_{3}^+)$
we wrote the mass matrix in the charged scalar sector,
in the basis $\Phi_E^0 = (\xi_d^0,\xi_u^0,{\tilde\nu}^0_+)$
we wrote the mass matrix in the CP-even neutral scalar sector, and
in the basis $\Phi_O^0 = (\phi^0_d,\phi^0_u,{\tilde\nu}^0_-)$
we wrote the mass matrix in the CP-odd neutral scalar sector.

The new charged scalar and CP-even and CP-odd neutral scalar 
mass-eigenstates are obtained
by diagonalizing the above-mentioned matrices. They are:
$\Phi_C = \left(H^+, G^+, \tilde\tau^+\right),  
\Phi_E = \left(H, h, \tilde\nu_+^\tau\right)$, and
$\Phi_O = \left(A, G^0, \tilde\nu_-^\tau\right)$. 
In the limit $b_3 \to 0$:
$H,h,A,H^+$ become the usual ones.
Rotating with the diagonalizing $R_{C,E,O}$ (for Charged, Even-CP,
Odd-CP) matrices, one goes from the $\Phi$s to the $\Phi^0$s. 
All depends on the four parameters
$A^0,~m_{s\nu}^0$ (the masses in the limit $b_3 \to 0,$),
$b_3$ and $\tan\beta$.

Let us sub-divide type B into two types according to the 
scalar and fermion in the loop:\\
\underline{Type B1:} Here  $S_\alpha=\Phi_{C,\alpha};~f_i=u_i$ 
with $\alpha=1,3;~i=1,2,3$.
The $a$ couplings are equal to their values in the SM.
The $b$ couplings include elements of the rotation
matrix (for the charged fields) $R_C$ and $\lambda^\prime$, 
and $g_Z^{\alpha\beta}=-e\cot{2\theta_W}\delta_{\alpha\beta}.$\\
\underline{Type B2:} In this case 
$S_\alpha=\Phi_{E,\alpha}$ and $\Phi_{O,\beta};~
f_i=d_i$ with $\alpha=1,2,3$, and $\beta=1,3;~ i=1,2,3$.
This is the only case for which our generic form is insufficient.
This fact results from the appearance of two new diagrams proportional
to a scalar-vector-vector coupling ($ZZ\Phi_E$ in our case). The
other eight diagrams are special cases of the generic ones in Fig. \ref{fig1}.

Inserting parameters consistent with the data we found
that for type B: ${\rm Br}(\zbs)\lsim 10^{-6}$.
  
\section{EXPERIMENTAL FEASIBILITY}

Let us briefly comment about the 
feasibility of observing 
(or limiting) a signal of $BR(Z \to b \bar s)\sim  
10^{-6}$, at a Linear Collider producing $10^9$ Z-bosons. 
Such a signal leads to 
one $b$-jet 
and one $q$-jet, where $q$ stands for quarks lighter than $b$.
The main background is from $Z\to b{\bar b}$. Using what,
we think, are realistic efficiencies we find that
a new physics signal $Z \to b \bar s$, with 
a branching ratio of order $10^{-6}$, can reach beyond
the 3-sigma level \cite{ourZbs}.
We can also get a clue about how low one can go in the value 
(or limit) of $BR(Z\to b \bar s)$ with $10^9$ Z-bosons, from the fact that 
the DELPHI preliminary result reached \cite{DELPHI}
$BR(Z\to b \bar s) < {\cal O}(10^{-3})$
with ${\cal O}(10^{6})$ $Z$-bosons.
Scaling this limit, especially with the expected advances in $b$-tagging 
and identification of non-$b$ jets methods, an ${\cal O}(10^{-6})$ branching 
ratios should be easily attained at a Giga-$Z$ factory. 
One needs realistic simulations as feasibility studies for this important
rare $Z$ decay mode.

\section{SUMMARY AND CONCLUSIONS}
 
Our results are best summarized in Table 1 which shows the best 
values for ${\rm Br}(\zbs)$ in
extensions of the SM discussed in this talk. The
SM result is given for comparison. Note that we have not included
interference with the SM as each of the values ``stands alone''.
In some cases such interference may increase the branching ratio
to $\sim 10^{-7}$. 
\begin{table}[hbt]
\caption{Maximal ${\rm Br}(\zbs)$ in extensions of the SM}  
\begin{tabular}{lll}
\hline
{\bf model}     & {\bf scalars in loop }& {\bf Br} \\ 
\hline 
SM              & $W^+$ (no scalars)    & $3\times 10^{-8}$ \\
2HDMII          & $H^+$                 & $10^{-10}$ \\
T2HDM           & $H^+$                 & $10^{-8}$ \\
$\sb$-$\ss$ mix & $\sb$-$\ss$ admix     & $10^{-6}$ \\
$\st$-$\sc$ mix & $\st$-$\sc$ admix     & $10^{-8}$  \\
tri-tri $\rpv$  & $\sq,\snu,\slep$      & $10^{-10}$  \\
tri-bi $\rpv$   & $\snu$-$h,H,A$ admix  &$10^{-6}$ \\
\hline
\end{tabular} 
\end{table}
 
We conclude that Giga-$Z$ experiments will have the opportunity to
place significant limits, or hopefully discover the scenario beyond
the SM, by searching for hadronic (and leptonic \cite{Illana})
neutral current flavor changing transitions.\\

\n {\it Acknowledgements}:\\
I would like to thank my collaborators, especially Shaouly Bar-Shalom,
for teaching me so much. I would also like to express my appreciation to
the organizers of the meeting for a job well done. Thanks also
to members of the theory group in DESY (Hamburg) who gave me the peace
of mind  I needed to prepare my talk. This research was supported in
part by the US-Israel Binational Science Foundation, by the Israel
Science Foundation and by the Fund for Promotion of Research at the
Technion.


\begin{thebibliography}{99}
\bibitem{ourZbs} D. Atwood, S. Bar-Shalom, G. Eilam and A. Soni,
Phys. Rev. D66 (2002) 093005. 

\bibitem{Gaill73}
M.K.~Gaillard and B.W.~Lee,
Phys.\ Rev.\ D10 (1974) 897.

\bibitem{Lunghi}
E.~Lunghi,
hep-ph/0210379.

\bibitem{toprare} B. Mele, hep-ph/0003064. 

\bibitem{mutoegam} 
Y.~Kuno and Y.~Okada,
Rev.\ Mod.\ Phys.\  {\bf 73} (2001) 151

\bibitem{GigaZ} J.A. Aguilar-Saavedra {\it et al.},
ECFA/DESY LC Physics Working Group, hep-ph/0106315.

\bibitem{Illana} J.I Illana, these proceedings.

\bibitem{SMzbs} 
G.~Mann and T.~Riemann,
Annalen Phys.\  {\bf 40} (1984) 334;
M. Clements {\it et al.}, Phys. Rev. {\bf D27} (1983) 570;
V. Ganapathi {\it et al.}, {\it ibid.} {\bf D27} 1983) 579.
 
\bibitem{DELPHI}
 J. Fuster, F. Martinez-Vidal and P. Tortosa,
preprint DELPHI 99-81 CONF 268, June 1999. 
The preprint can be downloaded 
from: http://documents.cern.ch/cgi-bin/setlink?base=preprint\&categ=cern\&id\\
=open-99-393. SLD plans to improve this limit; see: S. Walston's talk at 
DPF2002.

\bibitem{Pas:Velt} G. Passarino and M. Veltman, 
Nucl.\ Phys.\ {\bf B160} (1979) 151. See \cite{ourZbs}
for our notation.

\bibitem{our_rev}
D. Atwood, S. Bar-Shalom, G. Eilam and A.Soni, Phys. Rep. {\bf 347} (2001) 1.

\bibitem{our4}
B. Grzadkowski, J.F. Gunion and P. Krawczyk, 
Phys. Lett. {\bf B268} (1991) 106. 

\bibitem{our17:19}
K. Kiers, A. Soni and G.-H. Wu,
Phys.\ Rev.\ {\bf D59} (1999) 096001, {\it ibid.}
{\bf D62} (2000) 116004; 
G.-H. Wu and A. Soni, {\it ibid.}
{\bf D62} 056005 (2000).

\bibitem{our20} M. Misiak, S. Pokorski and J. Rosiek, 
Adv.\ Ser.\ Direct.\ High Energy Phys.\ {\bf 15} (1998) 796. 

\bibitem{our21} 
J.L. Diaz-Cruz, H.-J. He and C.-P. Yuan,
Phys. Lett. {\bf 530} (2002) 179.

\bibitem{our22} 
J. Rosiek, 
Phys.\ Rev.\ {\bf D41} (1990) 3464, and hep-ph/9511250 (erratum).
See also \cite{ourZbs}. 

\bibitem{our23}
T. Besmer, C. Greub and T. Hurth, Nucl.\ Phys.\ {\bf B609} (2001) 359.

\bibitem{our6} M. Chemtob and G. Moreau,
Phys. Rev. {\bf D59} (1999) 116012.

\end{thebibliography}
\end{document}